\documentclass[aps,pre,twocolumn,superscriptaddress,amsmath,amssymb,showpacs]{revtex4}
\usepackage{graphicx}

\begin{document}

\title{Stability of two-dimensional spatial solitons in nonlocal nonlinear media}

\author{S. Skupin}
\altaffiliation{Present address: D\'epartement de Physique Th\'eorique et Appliqu\'ee, CEA/DAM
Ile de France, B.P. 12, 91680 Bruy\`eres-le-Ch\^atel, France}
\email{stefan.skupin@cea.fr}
\affiliation{ Laser Physics Center,  Research School of Physical Sciences and Engineering,  Australian National University, Canberra, ACT 0200, Australia}

\author{O. Bang}
\affiliation{COM$\bullet$DTU Department of Communications, Optics \& Materials, Technical University of Denmark, DK-2800 Kongens Lyngby, Denmark}

\author{D. Edmundson}
\affiliation{ANU Supercomputer Facility, Australian National University, Canberra ACT 0200, Australia}

\author{W. Krolikowski}
\affiliation{ Laser Physics Center, Research School of Physical Sciences and Engineering,  Australian National University, Canberra, ACT 0200, Australia}

\date{\today}

\begin{abstract}We discuss existence and stability of two-dimensional solitons in media with spatially nonlocal nonlinear response. We show that such systems, which include thermal nonlinearity and dipolar Bose Einstein condensates, may support a variety of stationary localized structures - including rotating spatial solitons. We also demonstrate that the stability of these structures critically depends on the spatial profile of the nonlocal response function.
\end{abstract}


\maketitle
\section{Introduction}
A spatial optical soliton is a beam  which propagates in nonlinear medium without changing its structure. Formation of solitons is a result of a balance between diffraction caused by the finite size of the wave and the nonlinear changes of the refractive index  of the medium induced  by the wave itself \cite{StegSeg99,KivsharAgrawal}. When this balance can be maintained dynamically the soliton exists as a robust object withstanding even strong perturbation.  It appears that solitons, or solitary waves, are ubiquitous in nature and have been identified in many other systems such as plasma, matter waves or classical field theory.  

Solitons have been typically considered in the context of the so called local nonlinear media. In such media the refractive index change induced by an optical beam in a particular point depends solely on the beam intensity in this very point. However, in many optical systems the nonlinear response of the medium is actually a spatially nonlocal function of the wave intensity. This means that index change in particular point depends on the light intensity in a certain neighborhood of this point. This occurs, for instance, when the nonlinearity is associated with some sort of transport processes such as heat conduction in media with thermal response \cite{Litvak66,Litvak75,DF95}, diffusion of charge carriers \cite{Wright85,Ultanir} or atoms or molecules in atomic vapors \cite{Happer77,Suter93}. It is also the case in systems exhibiting a long-range interaction of constituent molecules or particles such as in nematic liquid crystals \cite{McLaughlin95,Assanto03,Conti04,Peccianti05} or dipolar Bose Einstein condensates \cite{ParSalRea98,Gries05,Cooper05,Zhang05,Pedri05}. Nonlocality is thus a feature of a large number of nonlinear systems leading to novel phenomena of a generic nature. For instance, it may promote modulational instability in self-defocusing media \cite{WK01,WK03,Wyller02,WK04}, as well as suppress wave collapse of multidimensional beams in self-focusing media \cite{Tur85,Bang02,DF95}. Nonlocal nonlinearity may even represent parametric wave mixing, both in spatial \cite{Nikola03} and spatio-temporal domain \cite{Larsen06} where it describes formation of the so called X-waves. Furthermore, nonlocality significantly affects soliton interaction leading to formation of bound state of otherwise repelling bright or dark solitons \cite{Fratalocchi04,Nikola04,Dreischuh06,Rasmussen05}. It has been also shown that nonlocal media may support formation of stable complex localized  structures. They include multihump \cite{Kolchugina80,Mironov81} and vortex ring solitons \cite{Briedis05,Yakimenko05}.

The robustness  of nonlocal  solitons has been attributed to the fact that by its nature,  nonlocality acts as  some sort of spatial averaging of the nonlinear response of the medium.  Therefore perturbations, which normally would grow quickly in a local medium, are being averaged out by nonlocality ensuring a greater stability domain of solitons \cite{Briedis05,Servando06}. However, it turns out that such an intuitive physical picture of nonlocality-mediated soliton stabilization which has been earlier confirmed in studies of modulational instability and beam collapse, has only limited validity. Recent works by Yakimenko \cite{Yakimenko05}, Xu \cite{Xu05} and Mihalache \cite{Mihalache06} demonstrated  that even a high degree of nonlocality may not guarantee existence of stable high order soliton structures.

In this work we investigate the propagation of finite beams in nonlocal nonlinear media. We demonstrate that while nonlocality does stabilize solitons, the stability domain is strongly affected by the actual form of the spatial nonlocality.

This paper is organized as follows: In Section \ref{sec_model} we introduce the mathematical models under consideration and the ansatz for the solutions we are interested in. Section \ref{sec_besselk} deals with spatial optical solitons in media with a thermal nonlinearity. A remarkably robust new type of rotational soliton is presented. In Section \ref{sec_bec} we treat the more complicated case of two-dimensional solitons in dipolar Bose Einstein condensates (BEC). Due to the mixture of local and nonlocal type of nonlinearity a variety of solutions can be stabilized. Finally, in Section \ref{remarks} we discuss the observed dynamics of the soliton structures. 

\section{\label{sec_model}Model Equations}

We consider physical systems governed by the two-dimensional nonlocal nonlinear Schr{\"o}dinger (NLS) equation
\begin{equation}
\label{NLS}
i \frac{\partial}{\partial z} \psi +  \left(\frac{\partial^2}{\partial x^2} + \frac{\partial^2}{\partial y^2} \right)\psi  + \theta \psi = 0.
\end{equation}
where $\theta$ represents the spatially nonlocal nonlinear response  of the medium. Its form depends on the details of a particular physical system. In the following  we will consider three nonlocal models. 

The first one is rather unphysical but extremely instructive, the so called Gaussian model of nonlocality. In this model $\theta$ is given as 
\begin{equation}
\label{NL_gaussian_int}
\theta = \frac{1}{2\pi}\iint\mathrm{e}^{-\frac{|\vec{r}-\vec{r}^{\prime}|^2}{2}} \left|\psi(\vec{r}^{\prime},z)\right|^2d^2\vec{r}^{\prime}.
\end{equation}
In spite of the fact that there is no known physical system which would be described by a Gaussian response, this model has served as a phenomenological example of a nonlocal medium, enabling, thanks to its form, an analytical treatment of the ensuing wave dynamics.  

The second model, referred to as thermal nonlinearity, describes, for instance, the effect of plasma heating on the propagation of electromagnetic waves \cite{Litvak66}. In this case  $\theta$ is governed by the following diffusion-type equation \cite{Litvak66,Litvak75}
\begin{equation}
\label{NL_thermal_diff}
\theta - \left(\frac{\partial^2}{\partial x^2} + \frac{\partial^2}{\partial y^2} \right)\theta = |\psi|^2,
\end{equation}
which is valid for the typical spatial diffusion scale large compared to the operating wavelength \cite{Litvak66,Dabby68,Yakimenko05}.  Solving formally Eq.\ (\ref{NL_thermal_diff})   in the Fourier space yields 
\begin{equation}
\label{NL_thermal_int}
\theta = \frac{1}{2\pi}\iint\mathfrak{K}_0(|\vec{r}-\vec{r}^{\prime}|) \left|\psi(\vec{r}^{\prime},z)\right|^2d^2\vec{r}^{\prime},
\end{equation}
where $\mathfrak{K}_0$ is the modified Bessel function of the second kind and $\vec{r}=x\vec{e}_x+y\vec{e}_y$ denotes the transverse coordinates. 

The third system considered here is the model of a dipolar Bose Einstein condensate where the nonlocal character of the interatomic potential is due to a  long-range interaction of dipoles \cite{Santos00}. Such condensate has been realized recently in experiments with Chromium atoms which exhibit strong magnetic dipole moment \cite{Gries05}. Considering only two transverse dimensions and time (which plays a role analogous to that of the propagation variable $z$) one arrives at the following formula~\cite{Pedri05}

\begin{equation}
\label{BEC_response}
\theta = \alpha|\psi|^2 + \frac{1}{2\pi}\iint\mathfrak{R}(|\vec{r}-\vec{r}^{\prime}|) \left|\psi(\vec{r}^{\prime},z)\right|^2d^2\vec{r}^{\prime},
\end{equation}
and
\begin{equation}
\label{NL_BEC}
\mathfrak{R} = \iint \frac{1-\sqrt{\pi}k\mathrm{e}^{k^2}\mathrm{erfc}(k)}{2\pi} \mathrm{e}^{i(k_xx+k_yy)}dk_xdk_y,
\end{equation}
with $k=\sqrt{k_x^2+k_y^2}$ and $\mathrm{erfc}$ being the complementary error function \cite{Giovanazzi02,Pedri05}. 
Interestingly, this model contains both, local and nonlocal components. The parameter $\alpha$ controls the sign as well as the relative strength of the local component of nonlinearity. This model is somewhat controversial and its validity has been recently questioned \cite{Konotop02}. However this issue is beyond the scope of this paper and we consider Eq.(\ref{BEC_response}) here in its own right, as another example of nonlocal nonlinearity. 

Note that the above three models are represented in dimensionless form. This means that only the spatial extent of the actual solution $\psi$ determines whether one operates in a "local" ($|\psi|^2$ varies slowly over the transverse lengths of the order unity) or "strongly nonlocal" ($|\psi|^2$ changes fast) regime. The "conventional" degree of nonlocality, the actual width of the response function, is scaled out.

\begin{figure}
\includegraphics[width=8cm]{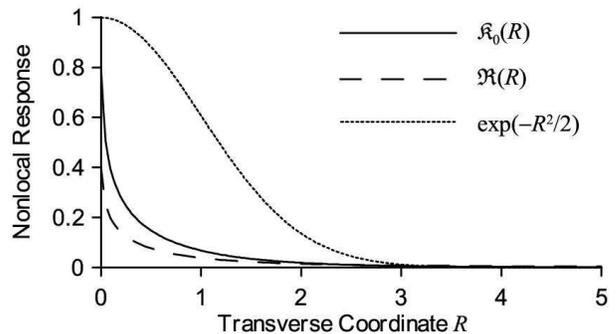}
\caption{\label{fig_response}Nonlocal response functions for thermal nonlinearities ($\mathfrak{K}_0$) and dipolar BEC's ($\mathfrak{R}$). For comparison, a Gaussian response function is shown, too.}
\end{figure}

It is known that the thermal model [Eq.\ (\ref{NL_thermal_diff})] permits, apart from the ground state, stable single-charge vortex solutions only, whereas multi-charge vortices appear to be  unstable \cite{Yakimenko05}. In contrast, when a Gaussian response function is used [Eq.(\ref{NL_gaussian_int})], also stable multi-charged vortices may exist \cite{Briedis05}. Figure \ref{fig_response}
illustrates  the nonlocal response functions for the two physically relevant  systems introduced above, and,  as a  comparison the (unphysical) Gaussian response. It is obvious that the thermal and condensate response functions exhibit qualitatively the same spatial character. Therefore, one can expect a similar behavior of solitary solutions for beams in media with thermal nonlinearities and matter waves in dipolar BEC's, especially with respect to their stability.   

In the following analysis we will be interested in two classes of nonlinear localized solutions: classical  solitons with stationary intensity and phase distribution, and a new recently introduced type of rotating solution, the so called "azimuthons" \cite{Desyatnikov05}. In case of the standard solitons we are looking for stationary solutions in the form
\begin{equation}
\label{soliton_ansatz}
\psi (r,\phi ,z) = U(r,\phi )\mathrm{e}^{i\lambda z},
\end{equation}
where $\lambda$ is the propagation constant. Here we changed from Cartesian to cylindrical coordinates for technical convenience.
In the highly nonlocal limit the soliton profile  $|U|^2(r,\phi)$ is expected to be very narrow compared to a width of the response function (which is a fixed quantity in our dimensionless system). Under this condition, for the Gaussian response the nonlinear term Eq.\ (\ref{NL_gaussian_int}) can be represented in simplified form \cite{Bang02}
\begin{equation}
\theta = \frac{1}{2\pi}P_0\exp(-r^2/2),
\end{equation}
where $P_0=\int|U|^2d^2\vec{r}$ is the total power. Now the nonlinear Schr\"odinger equation  Eq.\ (\ref{NLS}) becomes linear and local. It describes the evolution of an optical beam trapped in an
effective waveguide structure ("potential") with the profile given by the Gaussian nonlocal response function. This highly nonlocal limit was first explored  by  Snyder and Mitchell in the context of the 
"accessible solitons" \cite{Snyder97}. In our scaling the strongly nonlocal limit corresponds to high power  $P_0\rightarrow\infty$.  Large power $P_0$ means now a deep "potential" $-\theta$ in Eq.\ (\ref{NLS}) and stronger confinement of modes (solitons) in the vicinity of  $r=0$. We exploit this accessibility character of the solitons in our numerical scheme (see Appendix \ref{sec_numerics}). 

The second class of solutions we deal with, the azimuthons, are a straightforward generalization of the ansatz (\ref{soliton_ansatz}). They represent spatially rotating structures and hence involve an additional parameter, the rotation frequency $\Omega$ 
\begin{equation}
\label{azimuthon_ansatz}
\psi (r,\phi ,z) = U(r,\phi -\Omega z)\mathrm{e}^{i\lambda z}.
\end{equation}
For $\Omega=0$, azimuthons become ordinary (nonrotating) solitons. For example, the most simple family of azimuthons is the one connecting the dipole soliton with the single charged vortex soliton (for fixed propagation constant $\lambda$). The single charged vortex consists of two dipole-shaped structures in real and imaginary part of $U$ with equal amplitude. If these two amplitudes are different we have a rotating azimuthon, if one of the amplitudes is zero we have the dipole soliton. In the following we will refer to this amplitude ratio in terms of the modulation depth
\begin{equation}
\label{modulation_depth}
n = \frac{\left| \max \mathrm{Re} U - \max \mathrm{Im} U \right|}{\max \left| U \right|}.
\end{equation}
In case of a Gaussian nonlocal response function, for instance, the azimuthons arise naturally again via the fact that $\theta = P_0\exp(-r^2/2)/2\pi$ for high powers. Since Eq.\ (\ref{NLS}) becomes linear, solitons will converge to the linear modes of the corresponding "potential" $-\theta$. In this strongly nonlocal limit azimuthons, as introduced above, will converge to two degenerate dipole modes with the phase difference of $\pi/2$ and unequal amplitudes. As pointed out in \cite{Desyatnikov05}, the rotation frequency $\Omega$ is determined by the amplitude ratio  of the two degenerate modes (modulation depth) and, of course, the propagation constant $\lambda$.

\section{\label{sec_besselk}Beam propagation in media with a thermal nonlinearity}

\begin{figure}
\includegraphics[width=8cm]{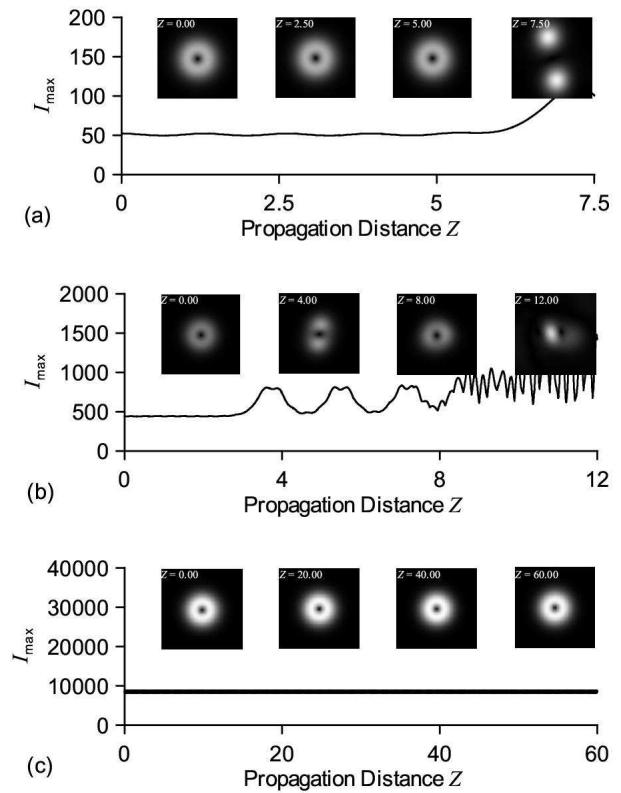}
\caption{\label{fig_bessel_vortex}Dynamics of the vortex states in thermal model: (a) unstable propagation at $P_0=200$, profiles are shown in a $xy$ box of about $5\times5$; (b) unstable propagation at $P_0=500$, profiles are shown in a $xy$ box of about $3\times3$; (c) stable at $P_0=2000$, profiles are shown in a $xy$ box of about $1.5\times1.5$.} 
\end{figure}

Let us first have a look at Eqs.\ (\ref{NLS}) and (\ref{NL_thermal_int}). It is known that in the local classical NLS, no stationary state exists. Finite beams either diffract if their power is too low or experience catastrophic collapse if their power exceeds a certain threshold. However, it has been already shown that in case of nonlocal NLS the catastrophic collapse is arrested and fundamental-type solitons can be stabilized. Moreover, it turns out that for sufficiently high degree of nonlocality (in our scaling this is equivalent to sufficiently high power), even  the  single charged vortex is reported to become stable \cite{Yakimenko05}. In fact, it is so far the only known stable stationary solution in this system, apart from the fundamental, single peak soliton. Indeed, Fig.\ \ref{fig_bessel_vortex}
shows that for power $P_0=2000$ the single charged vortex-state is stable, while for $P_0=200$ and $P_0=500$ it decays after a certain propagation distance. Of course, strictly speaking we cannot claim stability from numerical propagation experiments, since we can propagate beams over finite distances only. However, from our results we can at least infer that the potential growth rates of unstable internal modes are very small. For instance, the vortex state in Fig.\ \ref{fig_bessel_vortex}(c) was stable over more than $10^6$ numerical $z$-steps. The small periodic oscillations visible in some of the curves of numerical figures are due to the excitation of stable internal modes of the solution by perturbations of the initial data. In fact, since we use approximate solutions computed by the method described in Appendix \ref{sec_numerics}, we always deal with perturbed initial data.

Both vortices with $P_0=200$ and $P_0=500$ are unstable, but their decay dynamics are quite different. The one with lower power [Fig.\ \ref{fig_bessel_vortex}(a)] breaks up into two ground state solitons, which move away from the center once formed which is a typical instability scenario for local NLS. In contrast, for higher powers (or, conversely - stronger nonlocality) the vortex initially breaks-up and later fuses to a single  ground state solution emitting remnants of  its original structure [see \ref{fig_bessel_vortex}(b)]. This behavior can be explained by the  nonlocality-induced broad single waveguide (or attractive potential) which  prevents fragments from leaving (unlike the break-up in local medium).  

\begin{figure}
\includegraphics[width=8cm]{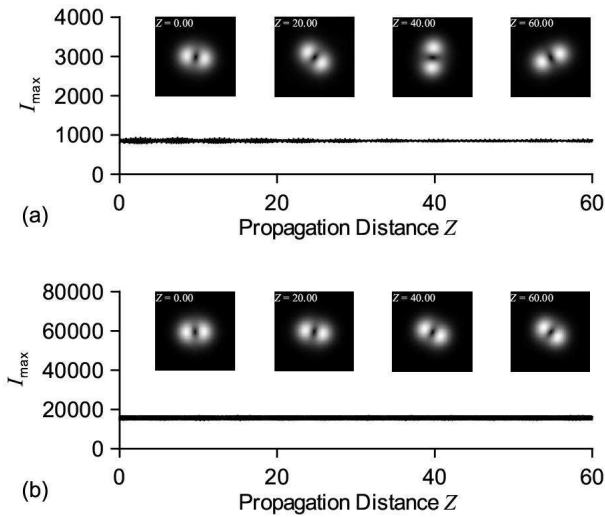}
\caption{\label{fig_bessel_azimuthon}Dynamics of azimuthons in thermal model of nonlinearity. Modulation depth $n=0.5$: (a) stable propagation at $P_0=500$, rotating with $\Omega\approx4.9$, profiles are shown in a $xy$ box of about $3\times3$; (b) stable propagation at $P_0=2000$, rotating with $\Omega\approx11$, profiles are shown in a $xy$ box of about $1.5\times1.5$.} 
\end{figure}

The decay behavior of the $P_0=500$ vortex indicates an effective power flux towards the center. Hence, one might expect that a rotation of the object could somehow balance this contraction. As mentioned in Sec.\ \ref{sec_model}, we are not looking at classical solitons only, but also at rotating azimuthons. At each power level $P_0$ we expect a family of azimuthons, allowing a continuous transition from the single charged vortex (modulation-depth $n=0$) to the flat-phased dipole-state (modulation-depth $n=1$). It turns out that the idea that the rotation of the azimuthon might  balance the contractive forces destroying the vortex is indeed  correct \cite{Servando06}. Fig.\ \ref{fig_bessel_azimuthon}(b)
shows that an azimuthon can be stable at power levels, where normally the vortex decays collapsing into a fundamental soliton ($P_0=500$). This shows  that azimuthons are more robust than vortices in this system. Hence, the azimuthons may be easier to realize than vortices in future attempts of experimental observation of higher order nonlinear beams in media with  thermal nonlinearity.

In our extensive numerical simulations of the thermal model  we were not able to observe any other stable, non-rotating dipole-states. Even at higher powers (stronger nonlocality) the simple dipole-state remained unstable. Moreover, we did not observe any decrease of the observed growth rates with increasing powers (shift of the break-up point to larger values of $z$). Also higher order states like quadrupole etc. were found to be always unstable in this system.

\section{\label{sec_bec}Two-dimensional dipolar Bose-Einstein condensates}

\begin{figure}
\includegraphics[width=8cm]{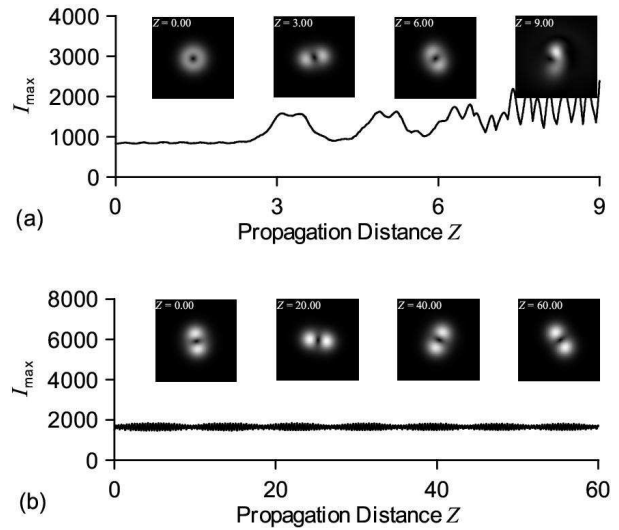}
\caption{\label{fig_bec_1000}Dynamics of localized structures in a BEC nonlinear model [Eq.(\ref{BEC_response})]. Power $P_0=1000$, $\alpha=0$: (a) unstable vortex-state; (b) stable azimuthon, rotating with $\Omega\approx5.1$, modulation-depth $n=0.5$. All profiles are shown in a $xy$ box of about $4\times4$} 
\end{figure}
\begin{figure}
\includegraphics[width=8cm]{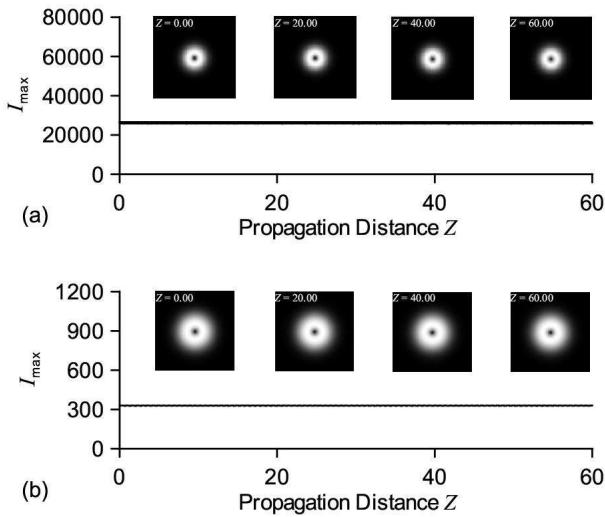}
\caption{\label{fig_bec_vortex}Stable vortex-states in the BEC model: (a) stable propagation at $P_0=5000$, $\alpha=0$, profiles are shown in a $xy$ box of about $2\times2$; (b) stable propagation at $P_0=1000$, $\alpha=-0.05$, profiles are shown in a  box of about $5\times5$.} 
\end{figure}

Compared to the previous system, the system of Eqs.\ (\ref{NLS}) and (\ref{BEC_response}) is more complicated. A free parameter $\alpha$ is involved, which controls both, the sign and the relative strength of an additional local contribution to the nonlinear response. Let us first consider the case $\alpha=0$. It turns out that in this case the system behaves qualitatively like the thermal model discussed in the preceding section. The azimuthons are found to be the most robust objects (apart from the stable ground state), and we observe the same stabilizing-by-rotation mechanism as above. The vortex at $P_0=1000$ shrinks to form a single ground state solution [Fig.\ \ref{fig_bec_1000}(a)], whereas the rotating azimuthon at the same power-level is stable due to its rotation [Fig.\ \ref{fig_bec_1000}(b)].
When we increase the power sufficiently the vortex-state also becomes stable [Fig.\ \ref{fig_bec_vortex}(a)], as expected.

\begin{figure}
\includegraphics[width=8cm]{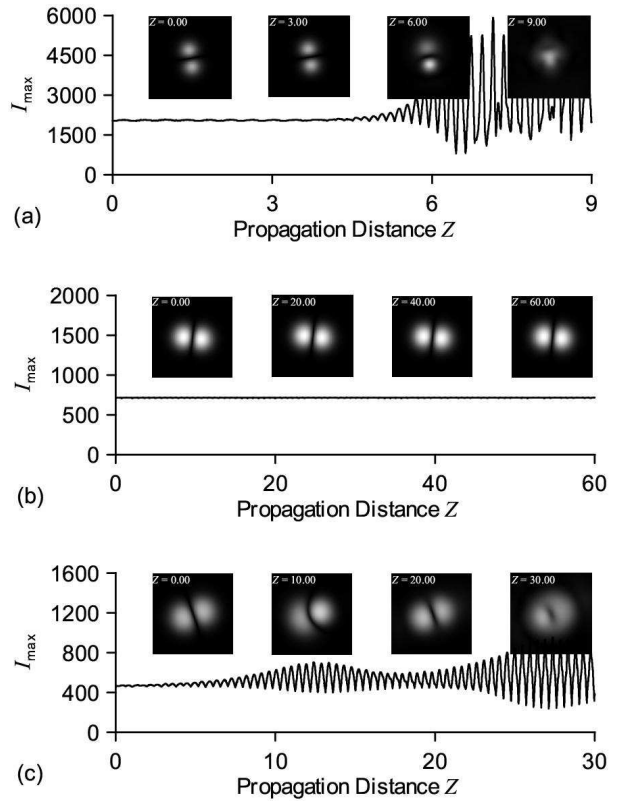}
\caption{\label{fig_bec_dipole}Dipole states of the BEC nonlocal model at $P_0=1000$: (a) unstable propagation, $\alpha=0$, profiles are shown in a $xy$ box of about $4\times4$; (b) stable propagation, $\alpha=-0.03$, profiles are shown in a $xy$ box of about $5\times5$. (c) unstable propagation, $\alpha=-0.05$, profiles are shown in a  box of about $5\times5$.} 
\end{figure}

The local part of the nonlinearity [Eq.(\ref{BEC_response})], can be either of focusing or defocusing nature, depending on the sign of the coefficient $\alpha$. If $\alpha$ is positive, stable solutions become unstable for $\alpha$ large enough. E.g., if we fix the power $P_0=1000$, the azimuthon with modulation-depth $n=0.5$ [stable for $\alpha=0$, see Fig.\ \ref{fig_bec_1000}(b)] turns out to be unstable already at $\alpha=0.01$. On the other hand, if $\alpha$ is negative, the otherwise unstable solutions can be stabilized. Fig.\ \ref{fig_bec_vortex}(b) shows the stable vortex-state at $P_0=1000$, $\alpha=-0.05$, but the corresponding vortex state at $\alpha=0$ is unstable [see Fig.\ \ref{fig_bec_1000}(a)]. Moreover, while playing with the parameter $\alpha$ we might even stabilize the dipole-state, as shown in Fig.\ \ref{fig_bec_dipole}(b).
This is remarkable, since the dipole seems to be always unstable if $\alpha=0$ [Fig.\ \ref{fig_bec_dipole}(a)], even for higher powers, and also in the system considered in Sec.\ \ref{sec_besselk}. However, the parameter $\alpha$ needs to be tuned carefully. If it is too large in absolute value, the stabilization mechanism fails [see Fig.\ \ref{fig_bec_dipole}(c)].

\begin{figure}
\includegraphics[width=8cm]{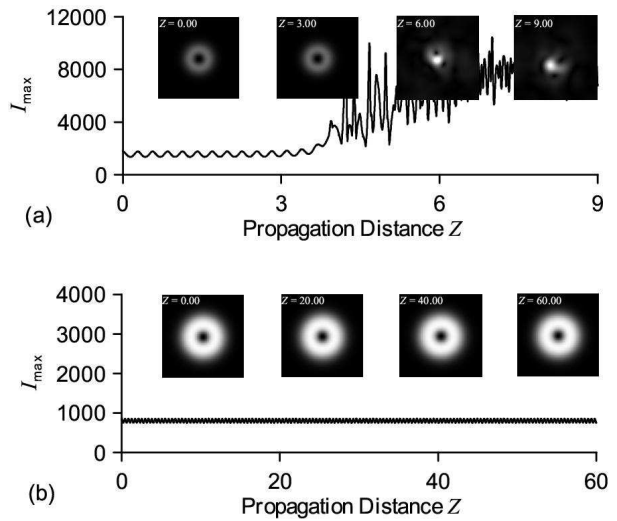}
\caption{\label{fig_bec_vortex2}Dynamics of double charge vortex states in a BEC nonlocal model with  $P_0=2000$: (a) unstable propagation, $\alpha=0$, profiles are shown in a $xy$ box of about $4\times4$; (b) stable propagation, $\alpha=-0.03$, profiles are shown in a $xy$ box of about $4\times4$.} 
\end{figure}

In a last example we show that even higher order solitons which are not related to the single-charged vortex (like the dipole) can be stabilized. The double charge vortex turns out to be stable with a small defocusing local contribution ($\alpha=-0.03$), as revealed by Fig. \ref{fig_bec_vortex2}.
The systematic studies of stability of other higher order solitons is beyond the scope of this paper. However, it seems that systems featuring a strong focusing nonlocal nonlinearity combined with a small local contribution of defocusing nature   are good candidates for observation of stable higher order solitons.

\section{\label{remarks}Discussion}

The examples considered above clearly demonstrate  that the stability of higher order nonlinear modes crucially depend on the shape of the nonlocal response function. Briedis \emph{et al.} \cite{Briedis05} suggested a very illustrative explanation why for a Gaussian response function higher order solitons can be stable. In the strongly nonlocal limit (high powers), the soliton shape $U$ becomes very narrow compared to the Gaussian response function. Hence, in this limit Eq.\ (\ref{NL_gaussian_int}) simplifies to $\theta = P_0\exp(-r^2/2)/2\pi$ where $P_0=\int|U|^2d^2\vec{r}$, and Eq.\ (\ref{NLS}) becomes linear and local. Hence we expect higher order states become stable at high enough powers in the Gaussian system. We may get at least some idea why in our physical systems [Eqs.\ (\ref{NL_thermal_int}) and (\ref{BEC_response})] higher order solitons, e.g. multi-charged vortices, are never stable: If we have a look at Fig.\ \ref{fig_response}, it is obvious that the   argument of Briedis \emph{et al.} \cite{Briedis05} breaks down for the response functions $\mathfrak{K}_0$ and $\mathfrak{R}$. The singularity of these functions at $r=0$ does not permit the simplification of the convolution integral like in the case for the Gaussian response function. No matter how  narrow the soliton shape $U$ may be, it will always feel the singularity of the response functions $\mathfrak{K}_0$ and $\mathfrak{R}$.

Even the role of the local nonlinearity (parameter $\alpha$) in the dipolar BEC [Eq.\ (\ref{BEC_response})] can be understood in this context. If $\alpha$ is positive, the focusing local contribution makes the singularity of the response-function [Eq.\ (\ref{BEC_response})] at $r=0$ "worse", and therefore tends to destabilize nonlinear solutions. On the other hand, for negative 
$\alpha$ the defocusing local contribution may balance the destabilizing effect of the nonlocal part of the response-function. 

A similar observation was reported recently by Xu \emph{et al.} \cite{Xu05} for the one-dimensional nonlocal NLS equation. In the 1D case, a thermal nonlinearity like Eq.\ (\ref{NL_thermal_int}) leads to an exponential response function $\exp(-2|x|)$, where $x$ is the transversal coordinate. In the strongly nonlocal regime for this exponential response only fundamental, dipole-, triple- and quadrupole-states are reported to be stable, whereas for the Gaussian response all higher order states turn out to become stable if the nonlocality is sufficiently high. Again, we see that due to the non-continuous first derivative of $\exp(-2|x|)$ the Briedis' {\em et al.} argument for stability does not hold. For completeness we investigated numerically  several (unphysical) response functions in 1D case. It seems that smooth response functions like $\exp(-x^2)$, $\mathrm{sech}(x)$, $1/[1+( x)^2]$ etc. permit stable multipole-states of high order in the strongly nonlocal limit. In contrast, response functions with an apex at $x=0$ like $\exp(-2|x|)$, $1/(1+2|x|)^2$, or $(n+1)(1-\sqrt[n]{|x|})/2$ with $|x|\le1$, $n\ge1$ allow only stable monopole-, dipole-, triple- and quadrupole-states in this regime. If we choose a response function with a singularity at $x=0$, like $(1/\sqrt{|x|}-1)/2$, at least also the quadrupole-state becomes unstable in the highly nonlocal regime. However, we can state that both in 2D and 1D, once the Briedis' {\em et al.} argument breaks down, even a strong nonlocality may not stabilize solitons of sufficiently high order.

\section{Conclusions}

In conclusion we discussed the propagation of two-dimensional solitons in nonlocal nonlinear media. We considered two physical relevant systems, optical beams in media with a thermal nonlinearity and two-dimensional dipolar Bose-Einstein condensates and compared them with the pure Gaussian nonlinear response which seems to support variety of stable solitons of high order. We demonstrated that while nonlocality does stabilize solitons, the stability domain is strongly affected by the actual form of the spatial nonlocality. We showed that both physically relevant systems support stable propagation of a new type of rotating solitons. In fact, our results suggest that these rotating structures are the most stable higher order nonlinear states in these systems, and might be even easier to realize experimentally than a single charge vortex or dipole solitons. Their appropriate input intensity and phase spatial profiles can be easily prepared using, for instance, spatial light modulator. We also showed that in a particular case of the BEC system, the mixture of local and nonlocal response stabilizes some higher order localized states in certain parameter regions.

\section{Acknowledgments}
This work was supported by the Australian Research Council. Numerical simulations were performed on the SGI Altix 3700 Bx2 cluster of the Australian Partnership for Advanced Computing (APAC) and on the IBM p690 cluster (JUMP) of the Forschungs-Zentrum in Jülich, Germany.

\appendix

\section{\label{sec_numerics}Numerical algorithm}

In order to find stationary localized structures supported by the nonlocal models Eqs.(\ref{NL_thermal_diff}),(\ref{BEC_response}, (\ref{NL_gaussian_int}) one must solve the 2-dimensional nonlinear Schr\"odinger equation with corresponding  nonlinear nonlocal term. 
In general, it is a quite difficult task to compute stationary solutions in 2D, especially when one cannot exploit certain symmetries, like, e.g., cylindrical symmetry as in the case of vortices. Here, we present a very fast and easy way to compute at least approximate solutions which relies on, the so-called, strongly nonlocal limit which we will illustrate using the Gaussian model. 

\begin{figure}
\includegraphics[width=8cm]{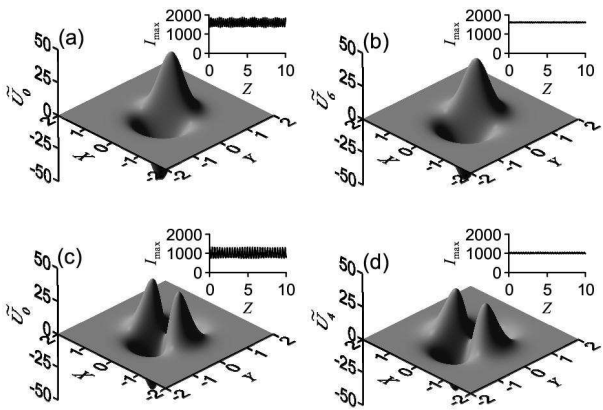}
\caption{\label{fig_gauss_sol} Approximate soliton solutions of Eqs.\ (\ref{NLS}) and (\ref{NL_gaussian_int}) for $P_0=1000$: (a) dipole state $\tilde{U}_0$ computed from $\tilde{\theta}_0 = P_0\exp(-r^2/2)/2\pi$; (b) dipole state $\tilde{U}_6$ after six iterations computed from $\tilde{\theta}_6$; (c) quadrupole state $\tilde{U}_0$ computed from $\tilde{\theta}_0 = P_0\exp(-r^2/2)/2\pi$; (d) quadrupole state $\tilde{U}_4$ after four iterations computed from $\tilde{\theta}_4$. The insets show the maximum intensity $I_{max}=\max_{x,y}|\psi|^2$ upon propagation, employing the approximate solution as an initial condition.}
\end{figure}

In the strongly nonlocal limit finding the exact stationary solution of the propagation equation reduces to the  eigenvalue problem $Av=\lambda v$, where vector  $v$ represents discretized  function $U(\vec{r})$ and $A$ is  a symmetric band matrix. These facts can be exploited to compute selected eigenvalues and eigenvectors iteratively (see \cite{Nikolai79,Parlett80,Demmel82,Golub96} for details). Then, using for instance,  $128\times128$ mesh it takes only a few seconds to obtain the desired solutions. Even if the condition $\theta = P_0\exp(-r^2/2)/2\pi$ is not exactly fulfilled, the obtained eigenvectors may be useful as approximate solutions. Moreover, it turns out that the following iterative process can be used to significantly improve the solution.  First, from the highly nonlocal approximation of the desired solution $\tilde{U}_0$ we compute improved form of the nonlocal term
\begin{equation}
\label{gaussian_iteration}
\tilde{\theta}_1 = \frac{1}{2\pi}\iint\mathrm{e}^{-\frac{|\vec{r}-\vec{r}^{\prime}|^2}{2}} \left|\tilde{U}_0(\vec{r}^{\prime})\right|^2d^2\vec{r}^{\prime},
\end{equation}
and use it to get a new eigenvalue problem $\tilde{A}_1 v=\lambda v$. By solving this new problem, we get $\tilde{U}_1$ which, again, can be used in the next iteration, and so on. There is no guarantee that $\tilde{U}_n$ converges to the exact solution $U$, and in general it does not. However,  by computing the residuum of Eq.\ (\ref{NLS}) with $U=\tilde{U}_n$ it is possible to monitor progress of  the iteration process. Fig.\ \ref{fig_gauss_sol}
\begin{figure}
\includegraphics[width=8cm]{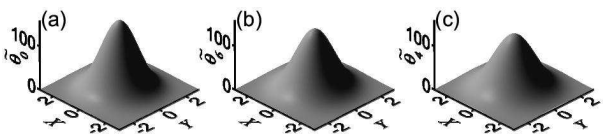}
\caption{\label{fig_gauss_theta} (a) $\tilde{\theta}_0 = P_0\exp(-r^2/2)/2\pi$ used to compute the approximate dipole- and quadrupole-states in Fig.\ \ref{fig_gauss_sol}(a) and \ref{fig_gauss_sol}(c); (b) $\tilde{\theta}_6$ used to compute the approximate dipole-state in Fig.\ \ref{fig_gauss_sol}(b); (c) $\tilde{\theta}_4$ used to compute the approximate quadrupole-state in Fig.\ \ref{fig_gauss_sol}(d).} 
\end{figure}
shows two examples of this technique, performed on the dipole- and quadrupole state at power $P_0=1000$. The actual $\tilde{\theta}$ used are shown in Fig.\ \ref{fig_gauss_theta}. In both cases, the initial guess $\tilde{U}_0$ [Figs.\ \ref{fig_gauss_sol}(a) and \ref{fig_gauss_sol}(c)] computed from $\tilde{\theta}_0 = P_0\exp(-r^2/2)/2\pi$ [see Fig.\ \ref{fig_gauss_theta}(a)]
can be improved by 6 and 4 iterations, respectively [Fig.\ \ref{fig_gauss_sol}(b) and \ref{fig_gauss_sol}(d)]. The actual test of the approximate solutions is provided by employing them as initial data and propagate over a certain $z$ distance. The results are presented in the respective insets of Fig.\ \ref{fig_gauss_sol} which show that soliton solutions are stable (at least over a propagation distance of $z=10$). Of course, since we treat a nonintegrable system these solitons are expected to possess  internal modes \cite{Rozanov04}. If we use an approximate solution as initial data, some of these internal modes become excited and cause oscillations of soliton amplitude. The better the approximation the weaker the oscillations. This fact is clearly visible when comparing the plots in Fig.\ \ref{fig_gauss_sol}(a) and \ref{fig_gauss_sol}(c), and \ref{fig_gauss_sol}(b) and \ref{fig_gauss_sol}(d) respectively.

These two examples show that it is possible to compute useful approximate solutions using the above iteration technique. The big advantage is its efficiency since the computations involved take only a couple of seconds on up-to-date computers. Moreover, it turns out that the technique can be used with non-Gaussian response functions, like Eqs.\ (\ref{NL_thermal_int}) and (\ref{BEC_response}), as well. In fact, all solutions presented in this paper were computed as described above.

\end{document}